\documentclass[10pt,superscriptaddress,prl,aps,twocolumn,nofootinbib,preprintnumbers]{revtex4}

\usepackage{float}
\usepackage{amsmath,amssymb}
\usepackage{soul}

\usepackage{graphicx}
\usepackage[caption=false]{subfig}

\usepackage{dcolumn}
\usepackage{bm}

\newcommand{\noplace}[1]{\affiliation{Void}}

\usepackage[colorlinks=true,linkcolor=blue,linktoc=all,citecolor=blue]{hyperref}
\usepackage[capitalise]{cleveref}
\usepackage{url}

\usepackage{breakurl}

\begin{document}

\title{A Neural-Network Extraction of\\Unpolarised Transverse-Momentum-Dependent Distributions \\ \vspace{0.2cm}
\normalsize{\textmd{The \textbf{MAP} (Multi-dimensional Analyses of Partonic distributions) Collaboration}}}

\author{Alessandro Bacchetta}
\thanks{E-mail: alessandro.bacchetta@unipv.it -- \href{https://orcid.org/0000-0002-8824-8355}{ORCID: 0000-0002-8824-8355}}
\affiliation{Dipartimento di Fisica, Universit\`a di Pavia, via Bassi 6, I-27100 Pavia, Italy}
\affiliation{INFN - Sezione di Pavia, via Bassi 6, I-27100 Pavia, Italy}

\author{Valerio Bertone}
\thanks{E-mail: valerio.bertone@cea.fr -- \href{https://orcid.org/0000-0003-0148-0272}{ORCID: 0000-0003-0148-0272}}
\affiliation{IRFU, CEA, Universit\'e Paris-Saclay, F-91191 Gif-sur-Yvette, France}

\author{Chiara Bissolotti}
\thanks{E-mail: cbissolotti@anl.gov -- \href{https://orcid.org/0000-0003-3061-0144}{ORCID: 0000-0003-3061-0144}}
\affiliation{Argonne National Laboratory, PHY Division, Lemont, IL, USA}

\author{Matteo Cerutti}
\thanks{E-mail: mcerutti@jlab.org -- \href{https://orcid.org/0000-0001-7238-5657}{ORCID: 0000-0001-7238-5657}}
\affiliation{Hampton University, Hampton, Virginia 23668, USA}
\affiliation{Jefferson Lab, Newport News, Virginia 23606, USA}

\author{Marco Radici}
\thanks{E-mail: marco.radici@pv.infn.it -- \href{https://orcid.org/0000-0002-4542-9797}{ORCID: 0000-0002-4542-9797}}
\affiliation{INFN - Sezione di Pavia, via Bassi 6, I-27100 Pavia, Italy}

\author{Simone Rodini}
\thanks{E-mail: simone.rodini@desy.de -- \href{https://orcid.org/0000-0002-8057-5597}{ORCID: 0000-0002-8057-5597}}
\affiliation{Deutsches Elektronen-Synchrotron DESY, Notkestr. 85, 22607 Hamburg, Germany}

\author{Lorenzo Rossi}
\thanks{E-mail: lorenzo.rossi3@unimi.it --\href{https://orcid.org/0000-0002-8326-3118}{ORCID: 0000-0002-8326-3118}}
\affiliation{Dipartimento di Fisica, Universit\`a di Milano, Via Celoria 16, 20133 Milan, Italy}
\affiliation{INFN, Sezione di Milano, Via Celoria 16, 20133 Milan, Italy}

\begin{abstract}
  We present the first extraction of transverse-momentum-dependent distributions
  of unpolarised quarks from experimental Drell-Yan data using neural networks
  to parametrise their nonperturbative part. We show that neural networks
  outperform traditional parametrisations providing a more accurate description
  of data. This work establishes the feasibility of using neural networks to
  explore the multi-dimensional partonic structure of hadrons and paves the way
  for more accurate determinations based on machine-learning techniques.
\end{abstract}

\preprint{DESY-25-022}
\preprint{JLAB-THY-25-4221}

\maketitle

\subsection{Introduction}\label{sec:introduction}

Transverse-momentum-dependent (TMD) distributions provide an important window to
investigate the partonic structure of hadrons, enabling a deeper understanding
of the three-dimensional dynamics of quarks and gluons within them.
TMDs consist of both perturbative and nonperturbative components. While the
perturbative part can be calculated from first principles using Quantum
Chromodynamics (QCD), the nonperturbative part, which encapsulates long-distance
physics, by definition cannot be evaluated using perturbative methods and must
be extracted from experimental data. An accurate parametrisation of the
nonperturbative contribution to TMDs is therefore essential for reliable
extractions. 
The abundance of experimental measurements and the development of a robust
theoretical framework have favoured a remarkable progress of TMD studies in
recent years. Indeed, accurate phenomenological extractions for unpolarised
quark TMDs in the proton are now
available~\cite{Bacchetta:2017gcc,Scimemi:2017etj,Bertone:2019nxa,Scimemi:2019cmh,Bacchetta:2019sam,Bury:2022czx,Bacchetta:2022awv,Moos:2023yfa,Bacchetta:2024qre}.

Despite this active research landscape, explorations of different
parametrisations of the TMD nonperturbative part were limited to models based on
a small set of functions, such as exponentials and (weighted) Gaussians. While
these parametrisations are effective, they carry a significant bias that may
limit the accuracy of the models. This rigidity can hinder the ability to fully
capture the underlying physics and extract information from experimental data.
To overcome these limitations, a promising alternative is given by neural
networks (NNs). 

In this work, we present the first extraction of the unpolarised quark TMD
parton distribution functions (PDFs) in the proton using NNs to parametrise
their nonperturbative component, achieving next-to-next-to-next-to-leading
logarithmic (N$^3$LL) accuracy. This constitutes a proof-of-concept extraction,
as it is based on Drell-Yan (DY) production data only and neglects flavour
dependence of the intrinsic transverse momentum of quarks. Nevertheless, it
demonstrates that NN-based TMD extractions outperform traditional
nonperturbative parametrisations. Indeed, we provide evidence that current
experimental data encode complexities beyond the reach of traditional
parametrisations. We remark that NNs have been used in the past to study other
partonic distributions (see, \textit{e.g.},
Refs.~\cite{Cuic:2020iwt,AbdulKhalek:2022laj,Fernando:2023obn,Bertone:2024taw,NNPDF:2024nan}),
but never unpolarised TMDs. Our work paves the way for improved TMD extractions,
leveraging the flexibility of NNs to extract information on the multidimensional
structure of hadrons using data from current and future experiments.

\subsection{Formalism and parametrisation}\label{sec:parametrization}

We consider the DY process $h_A + h_B \rightarrow \ell^+ + \ell^- + X$ where two
hadrons with mass $M$ and four-momenta $P_A$ and $P_B$ collide with
center-of-mass energy squared $s=(P_A+P_B)^2$, and inclusively produce a lepton
pair with total four-momentum $q$ and invariant mass $Q\gg M$. If the transverse
momentum component $\bm{q}_T$ with respect to the collision axis satisfies the
condition $|\bm{q}_T| \equiv q_T \ll Q$, the differential cross section can be
written as
\begin{equation}
  \begin{split}
\label{e:DYZ_xsec}
&\frac{d\sigma^{\text{DY}}}{d q_T\, dy\, dQ}  = \frac{8 \pi \alpha^2 q_T}{9 Q^3}\, {\cal P}\, x_A\, x_B \, {\cal H}^{\text{DY}}(Q,\mu)\, \sum_a c_a(Q^2)
  \\
&\times \int_0^{\infty}\!\! db_T \, b_T J_0\big( b_T q_T \big) \hat{f}_1^a (x_A, b_T^2; \mu, \zeta_A) \, \hat{f}_1^{\bar{a}} (x_B,  b_T^2; \mu, \zeta_B),
  \end{split}
\end{equation}
where $y=\ln\sqrt{(q_{0}+q_{z})/(q_{0}-q_{z})}$ is the lepton-pair rapidity,
$\alpha$ is the electromagnetic coupling, ${\cal P}$ is a phase-space-reduction
factor accounting for possible lepton cuts,\footnote{See Appendix~C of
Ref.~\cite{Bacchetta:2019sam} for details.} $x_{A,B} = Q e^{\pm y}/\sqrt{s}$ are
the longitudinal momentum fractions carried by the incoming quarks, ${\cal
H}^{\text{DY}}$ is a perturbative hard factor encoding the virtual part of the
scattering, and the sum runs over all active quark flavours $a$ with $c_{a}$ the
quark electroweak charges.

In Eq.~\eqref{e:DYZ_xsec}, $\hat{f}^a_{1}$ is the Fourier transform of the
unpolarised TMD PDF of quark flavour $a$. It depends on the quark longitudinal
momentum fraction $x$ and on the variable $b_T=|\bm{b}_T|$, where $\bm{b}_T$ is
Fourier-conjugated to the quark intrinsic transverse momentum $\bm{k}_\perp$. It
also depends on the renormalisation scale $\mu$ and on the rapidity scale
$\zeta$ (with the constraint $\zeta_A \zeta_B = Q^4$). Such dependence arises
from the removal of ultraviolet and rapidity divergences~\cite{Collins:2011zzd}
and is controlled by corresponding evolution equations. The complete set of
equations (omitting unessential variables and indices) is given by
\begin{align}
  \begin{split}
\label{e:evoleq}
\frac{\partial \hat{f}_{1}}{\partial\ln\mu} &= \gamma(\mu,\zeta)\,,\quad \frac{\partial \hat{f}_{1}}{\partial\ln\sqrt{\zeta}} = K(\mu)\,, \\
\frac{\partial K}{\partial\ln\mu} &=  \frac{\partial
  \gamma}{\partial\ln\sqrt{\zeta}} = -\gamma_{K}(\alpha_{s}(\mu))\,,
\end{split}
\end{align}
where $\gamma$ and $K$ are the anomalous dimensions of renormalisation group and
of Collins--Soper equations, respectively, and $\gamma_{K}$ is the so-called cusp
anomalous dimension which relates the cross derivatives of $\hat{f}_{1}$.

Given a set of initial scales ($\mu_i,\zeta_i$), the solution to these
differential equations allows us to determine the $\hat{f}_{1}$ at any final
scales ($\mu_f,\zeta_f$). In addition, in the region of small transverse
separations $b_T$, the TMD PDF $\hat{f}_{1}$ can be matched onto unpolarized \textit{collinear} PDFs $f_1$ through a convolution with
perturbatively calculable matching coefficients $C$.

The resulting expression for the TMD PDF at the final scales ($\mu_f,\zeta_f$)
is
\begin{align}
  \begin{split}
    \label{e:evolved_TMDs}
\hat{f}&_{1}(x, b_T; \mu_f, \zeta_f) =
\big[C\otimes f_1\big] (x, b_T; \mu_i, \zeta_i)
\\
& \times \exp\bigg\{ K(\mu_{i})\ln\frac{\sqrt{\zeta_{f}}}{\sqrt{\zeta_{i}}}\\
& \quad +\int_{\mu_i}^{\mu_f} \frac{d\mu}{\mu}  \bigg[\gamma_{F}(\alpha_{s}(\mu))-\gamma_{K}(\alpha_{s}(\mu))\ln\frac{\sqrt{\zeta_{f}}}{\mu}\bigg]\bigg\} \; ,
  \end{split}
\end{align}
where $\gamma_F(\alpha_s(\mu)) = \gamma(\mu,\mu^2)$ and $\otimes$ indicates the
Mellin convolution over the longitudinal momentum fraction $x$. A convenient
choice for the initial scales is $\mu_i=\sqrt{\zeta_i} \equiv \mu_b =
2e^{-\gamma_{\rm E}}/ b_T$, with $\gamma_{\rm E}$ the Euler constant, in that it
avoids the insurgence of large logarithms in the anomalous dimension $K$ and in
the matching coefficients $C$.

The TMD PDF in Eq.~\eqref{e:evolved_TMDs} includes the resummation of large
logarithms of $b_T$ to all orders in perturbation theory. A given logarithmic
accuracy implies that each ingredient in Eq.~\eqref{e:evolved_TMDs} must be
computed to the appropriate perturbative accuracy. The present extraction
incorporates all the necessary ingredients to reach N$^{3}$LL
accuracy~\cite{Bertone:2019nxa}.

The introduction of the scale $\mu_{b}\sim 1/b_T$ requires a prescription to
avoid integrating in Eq.~\eqref{e:DYZ_xsec} over the QCD Landau pole
($\Lambda_{\rm QCD}$) in the large-$b_T$ region. To this purpose, we adopt the
same choice of
Refs.~\cite{Bacchetta:2017gcc,Bacchetta:2022awv,Bacchetta:2024qre,Bacchetta:2019sam,Cerutti:2022lmb}
and replace $\mu_{b}$ with $\mu_{b_*} = 2e^{-\gamma_{\rm E}}/b_*$, where
\begin{equation}
\label{e:bTstar}
b_*(b_T, b_{\text{min}}, b_{\text{max}}) = b_{\text{max}}  \bigg( \frac{1 - e^{ -b_T^4 / b_{\text{max}}^4 }}{1 - e^{ -b_T^4 / b_{\text{min}}^4 }} \bigg)^{1/4}   ,
\end{equation}
with
\begin{align}
b_{\text{max}} = 2 e^{-\gamma_{\rm E}}  \text{ GeV}^{-1} \quad
b_{\text{min}} = 2 e^{-\gamma_{\rm E}}/\mu_f \, .
\label{e:bminmax}
\end{align}
This choice guarantees that the variable $b_{*}$ rapidly saturates to
$b_{\text{max}}$ at large values of $b_T$, preventing $\mu_{b_*}$ from reaching
$\Lambda_{\rm QCD}$. However, $b_{*}$ also introduces spurious power corrections
that scale like
$(\Lambda_{\text{QCD}}/q_T)^k$~\cite{Catani:1996yz,Kulesza:2002rh,Laenen:2000de,Kulesza:2003wn},
with $k>0$. In the region $q_T \simeq \Lambda_{\text{QCD}}$, these power
corrections become sizeable and can be modelled by including in
Eq.~\eqref{e:evolved_TMDs} the nonperturbative function $f_{\text{\tiny NP}}$ as
follows:
\begin{equation}
  \begin{split}
\hat{f}&_{1}(x, b_T;  \mu_f, \zeta_f) =
\big[C\otimes f_1\big] (x, b_T; \mu_{b_*}, \mu_{b_*}^2)  \\
& \times \exp\bigg\{ K(b_*, \mu_{b_*}) \ln\frac{\sqrt{\zeta_{f}}}{\mu_{b_*}}\\
& \quad +\int_{\mu_{b_*}}^{\mu_f} \frac{d\mu}{\mu}  \bigg[\gamma_{F}(\alpha_{s}(\mu))-\gamma_{K}(\alpha_{s}(\mu))\ln\frac{\sqrt{\zeta_{f}}}{\mu}\bigg]\bigg\}\\
& \quad \quad \times f_{\text{\tiny NP}} (x, b_T; \zeta_f) \,.
\label{e:evolved_TMDs_b*}
  \end{split}
  \end{equation}
The nonperturbative function must satisfy the condition $f_{\text{\tiny NP}} \to
1$ for $b_T \to 0$ in order to recover the perturbative regime. It must also
grant that the TMD PDF is suppressed for large values of $b_T$ and $\zeta_f$. We
parametrise $f_{\text{\tiny NP}}$ using a NN but enforcing these physically
required constraints. We explored several different NN parametrisations and will
report on them in a future work. As a proof of concept, in this work we focus on
the following model:
\begin{equation}
f_{\text{\tiny NP}}(x, b_T; \zeta) = \frac{\mathbb{NN}(x, b_T)}{\mathbb{NN}(x,0)} \exp\left[ -g_2^2 b_T^2 \log \left(\frac{\zeta}{Q_0^2} \right) \right] \,,
\label{e:NNmodel}
\end{equation}
where, as customary, we split $f_{\text{\tiny NP}}$ into an ``intrinsic''
nonperturbative part, entirely parametrised by the NN (denoted as
$\mathbb{NN}$), and the nonperturbative contribution to the rapidity evolution,
encoded in the exponential function. The NN is taken with architecture
$[2,10,1]$, \textit{i.e.} with two inputs corresponding to $x$ and $b_T$, $10$
hidden nodes, and one output node. The activation function associated to the
nodes of the hidden layer is
\begin{equation}
\sigma(z) = \frac{1}{2} \left( 1+ \frac{z}{1 + |z|}  \right) \,,
\label{e:sigmoid}
\end{equation}
which resembles the more traditional sigmoid function but offers a significant
reduction of computational burden while granting an excellent quality of the
final result. The activation function for the outer layer is instead chosen to
be quadratic. The reference scale for the rapidity evolution is set to $Q_0 =
1$~GeV. The parametrisation of the function $f_{\text{\tiny NP}}$ in
Eq.~(\ref{e:NNmodel}) is engineered to match the constraints mentioned above,
namely $f_{\text{\tiny NP}} \to 1$ for $b_T \to 0$ and $f_{\text{\tiny NP}} \ll
1$ for large $b_T$ and $\zeta_f$. With this setup, we have a total of $42$ free
parameters, 41 associated to the NN and one ($g_2$) to the evolution.

The values of the best-fit parameters are obtained by minimising a $\chi^2$ that
accounts for all sources of experimental uncertainties. The minimisation is
performed using the Levenberg--Marquardt algorithm as implemented in the {\tt
Ceres-Solver} package~\cite{Agarwal_Ceres_Solver_2022}. An important aspect of
our analysis is that the gradient of the $\chi^2$ with respect to the free
parameters is evaluated \textit{analytically} by exploiting the ability to
compute the derivatives of the NN with respect to its parameters in a closed
form~\cite{AbdulKhalek:2020uza}. This feature is crucial to ensure a fast and
stable convergence of the minimisation procedure.

Finally, overfitting is a well known problem of phenomenological analyses based
on NNs~\cite{NNPDF:2017mvq, NNPDF:2021njg}. We avoid it by using the
cross-validation method~\cite{DelDebbio:2007ee}. Specifically, the data set is
split into two subsets: one for training and one for validation. The training
set is used to determine the best-fit parameters, while the validation set is
used to monitor the quality of the fit. The best-fit set of parameters is
determined by requiring the $\chi^2$ of the validation set to be minimal. In our
analysis, we divided the data set into validation and training sets of the same
size.

\subsection{Results}\label{sec:results}

In this section, we discuss the results for the fit of $f_{\text{\tiny NP}}$ in
Eq.~(\ref{e:NNmodel}) to the DY experimental data included in the most recent
analyses of the MAP Collaboration (see
Refs.~\cite{Bacchetta:2022awv,Bacchetta:2024qre} for more details). We consider
fixed-target data from Fermilab (E605~\cite{Moreno:1990sf},
E288~\cite{Ito:1980ev}, and E772~\cite{E772:1994cpf}) and collider data from
Tevatron (CDF~\cite{Affolder:1999jh,Aaltonen:2012fi},
D0~\cite{Abbott:1999wk,Abazov:2007ac,Abazov:2010kn}), RHIC
(STAR~\cite{STAR:2023jwh}), and the LHC
(LHCb~\cite{Aaij:2015gna,Aaij:2015zlq,Aaij:2016mgv},
CMS~\cite{Chatrchyan:2011wt,Khachatryan:2016nbe,CMS:2019raw},
ATLAS~\cite{Aad:2014xaa,Aad:2015auj,ATLAS:2019zci}).

The collinear PDFs $f_1$ in Eq.~\eqref{e:evolved_TMDs_b*} are taken from the
MSHT2020 set~\cite{Bailey:2020ooq} of the {\tt LHAPDF}
library~\cite{Buckley:2014ana} at next-to-next-to-leading-order, which is
necessary to achieve N$^3$LL. The strong coupling $\alpha_s$ is obtained from
the same PDF set. We propagate the uncertainties of collinear PDFs into TMD PDFs
as in Refs.~\cite{Bacchetta:2022awv,Cerutti:2022lmb}. In order to ensure
applicability of TMD factorisation, we impose the kinematic cut $q_T/Q < 0.2$.
We exclude all experimental data in the energy range of the $\Upsilon$ resonance
(9~GeV $< Q < 11$~GeV). Moreover, we neglect the PHENIX data of
Ref.~\cite{PHENIX:2018dwt}, originally included in the analyses of
Refs.~\cite{Bacchetta:2022awv,Bacchetta:2024qre}, because only two data points
survive the $q_T/Q$ cut and their description is typically poor for any
parametrisation of $f_{\text{\tiny NP}}$.

The propagation of the experimental uncertainties into the TMD PDFs is achieved
through Monte Carlo (MC) sampling: an ensemble of $N_{\text{rep}}=250$
fluctuations (replicas) of the experimental data set is generated accounting for
correlated uncertainties, and each replica is used to extract $f_{\text{\tiny
NP}}$.

In order to estimate the performance of our NN-based fit, we performed an
additional fit with the same settings (data set, perturbative order, etc.) but
parametrising $f_{\text{\tiny NP}}$ with the functional form used in
Ref.~\cite{Bacchetta:2022awv}, which features 12 free parameters. In
Tab.~\ref{t:chi2}, we compare the quality of the NN-based fit with this latter
fit referred to as MAP22 (see Ref.~\cite{MAPfits} for the full results). For
each data subset (fixed-target, RHIC, Tevatron, ATLAS, CMS, and LHCb) we list
the number of points included in the fit ($N_{\text{dat}}$) and the reduced
$\overline{\chi}^2$ ($=\chi^2/N_{\text{dat}}$) of the central replica,
\textit{i.e.} the fit to the experimental central values without MC
fluctuations. The total $\overline{\chi}^2$ is given in the  bottom line. For
each $\overline{\chi}^2$ value, we also provide separately uncorrelated
($\overline{\chi}_{D}^2$) and correlated ($\overline{\chi}_{\lambda}^2$)
contributions (see Appendix~B of Ref.~\cite{Bertone:2019nxa} for more details).

\begin{table}[h]
\centering\resizebox{\columnwidth}{!}{
\begin{tabular}{llcc}
\hline
\vphantom{\Big|}Experiment\quad & $N_{\text{dat}}$ & \multicolumn{2}{c}{$\overline{\chi}^2\;(\overline{\chi}_{D}^2+\overline{\chi}_{\lambda}^2)$} \\
\vphantom{\Big|}& & NN & MAP22 \\ \hline
\vphantom{\Big|} Fixed-target	  & 233 & $1.08  \; ( 0.98 + 0.10 )$   \quad & $0.91 \; ( 0.70 + 0.21 )$ \\
\vphantom{\Big|} RHIC	          & 7   & $1.11  \; ( 1.03 + 0.07 )$   \quad & $1.45 \; ( 1.37 + 0.08 )$ \\
\vphantom{\Big|} Tevatron	      & 71  & $0.80  \; ( 0.73 + 0.06 )$   \quad & $1.20 \; ( 1.17 + 0.04 )$ \\
\vphantom{\Big|} LHCb	          & 21  & $0.98  \; ( 0.88 + 0.10 )$   \quad & $1.25 \; ( 1.05 + 0.20 )$ \\
\vphantom{\Big|} CMS	          & 78  & $0.40  \; ( 0.38 + 0.02 )$   \quad & $0.41 \; ( 0.35 + 0.06 )$ \\
\vphantom{\Big|} ATLAS	          & 72  & $1.38  \; ( 1.09 + 0.29 )$   \quad & $3.51 \; ( 3.03 + 0.49 )$ \\ \hline
\vphantom{\Big|} Total	          & 482 & $0.97  \; ( 0.86 + 0.11 )$   \quad & $1.28 \; ( 1.09 + 0.20 )$ \\
\end{tabular}}

\caption{Breakdown of the reduced $\overline{\chi}^2=\chi^2/N_{\text{dat}}$ for
each subset included in the fit and for the total data set. Results obtained
with the parametrisation in Eq.~(\ref{e:NNmodel}) (NN) and that of
Ref.~\cite{Bacchetta:2022awv} (MAP22) are shown. $\overline{\chi}_{D}^2$ and
$\overline{\chi}_{\lambda}^2$ correspond to uncorrelated and correlated
contributions to $\overline{\chi}^2$, respectively~\cite{Bertone:2019nxa}.}
\label{t:chi2}
\end{table}

It is evident that the NN fit achieves a better description of data than MAP22,
not only at the level of the global $\overline{\chi}^2$ (0.97 for NN vs. 1.28
for MAP22), but also for almost all single subsets (with the only exception of
fixed-target). Particularly significant is the improvement for ATLAS, for which
the $\overline{\chi}^2$ value drops from 3.51 for MAP22 to 1.38 for NN. As the
ATLAS measurements are the most precise ones, this is a clear indication that
the NN parametrisation can better capture the information encoded in the data.
We also note that the correlated contributions $\overline{\chi}_{\lambda}^2$ for
the NN fit are generally smaller than for MAP22, \textit{i.e.} the NN fit is
able to describe the data without relying on large correlated shifts.

\begin{figure}[tbh]
    \centering
    \includegraphics[width=0.49\textwidth]{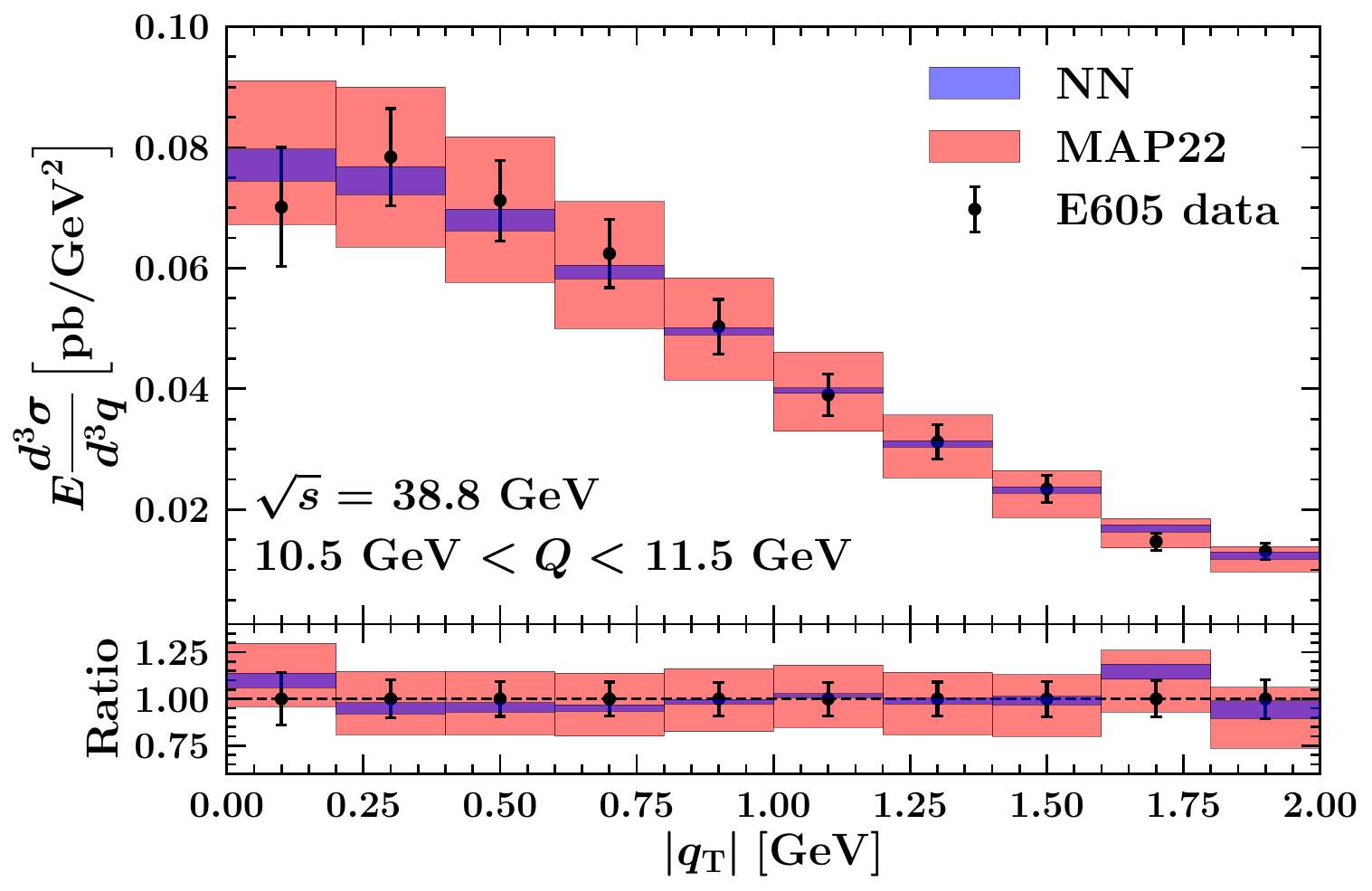}
    \includegraphics[width=0.475\textwidth]{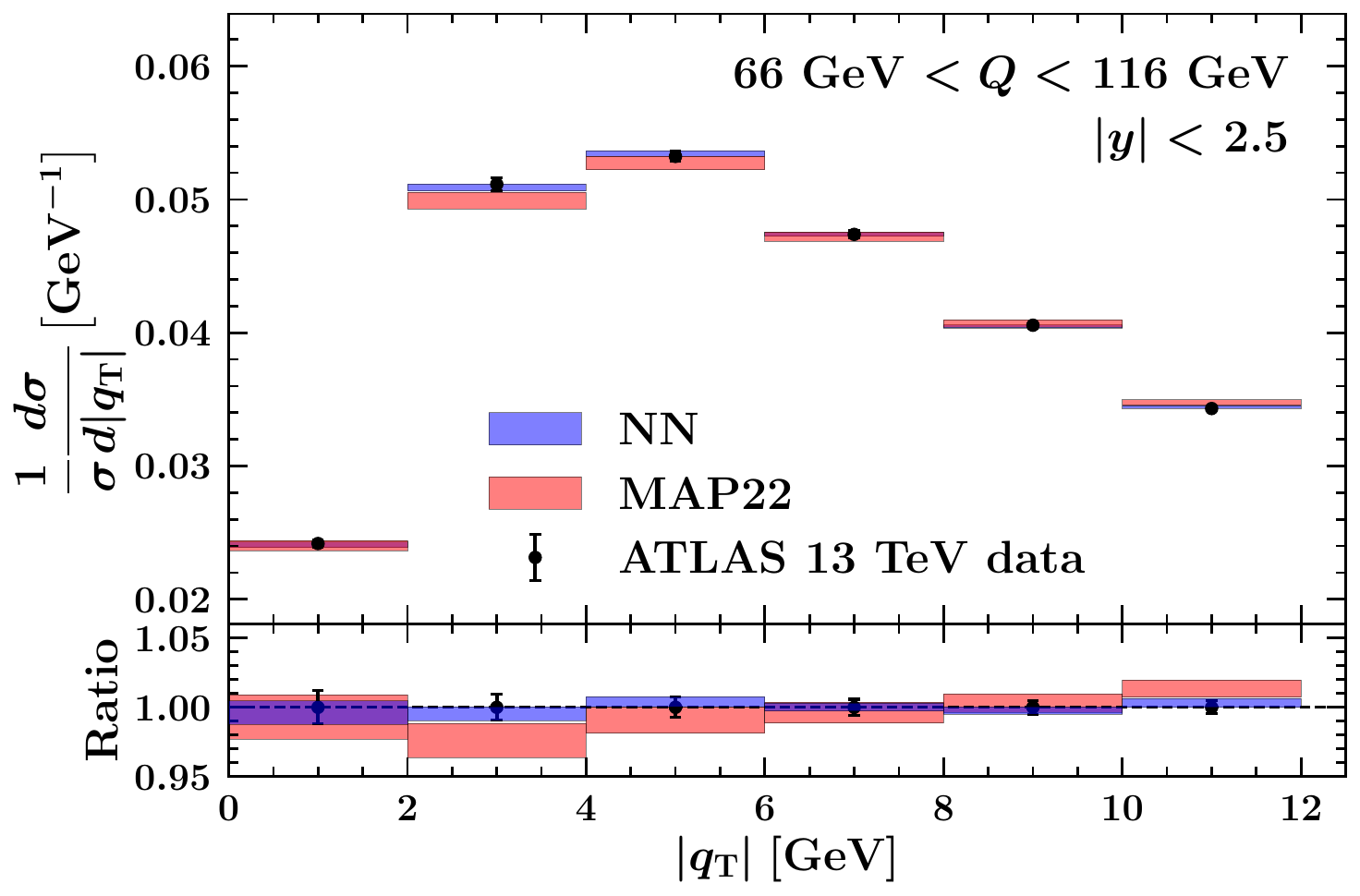}
    \caption{Comparison between experimental data (black dots) and results
    obtained with NN (blue band) and MAP22 (red band) fits. The top plot
    displays the $10.5~\mbox{GeV}<Q<11.5~\mbox{GeV}$ bin of the E605 data set,
    while the bottom plot displays the ATLAS measurements at 13 TeV. For each
    plot, upper and lower panels show the actual distributions and their ratios
    to the experimental central values, respectively. Theoretical uncertainty
    bands correspond to one-$\sigma$ uncertainties, error bars on experimental
    data display uncorrelated uncertainties only.}
    \label{f:obs_comp}
\end{figure}
A visual representation of this statement is given in Fig.~\ref{f:obs_comp},
where we show a comparison between experimental data and results of the fit for
a representative selection of data: a $Q$-bin from the fixed-target E605
experiment (top plot) and the ATLAS measurements at 13 TeV (bottom plot). Blue
and red bands correspond to one-$\sigma$ uncertainties of NN and MAP22 fits,
respectively, while experimental data points are shown as black dots along with
their uncorrelated uncertainties. The upper panel of each plot displays the
absolute distributions while the lower panel displays the distributions
normalised to the experimental central values.

In Fig.~\ref{f:obs_comp}, it is evident that uncertainty bands are significantly
different between NN and MAP22 fits, with the former being generally smaller
than the latter. This is a direct consequence of the larger systematic shifts
that affect MAP22 (see Tab.~\ref{t:chi2}). This is especially evident for the
E605 data where correlated uncertainties are particularly large. For the ATLAS
data, the size of the bands is comparable because systematic shifts are bound to
be small due to the small size of experimental uncertainties. We also observe
that the NN fit tends to better reproduce the shape of the ATLAS data
distribution.

\begin{figure}[tbh]
    \centering
    \includegraphics[width=0.49\textwidth]{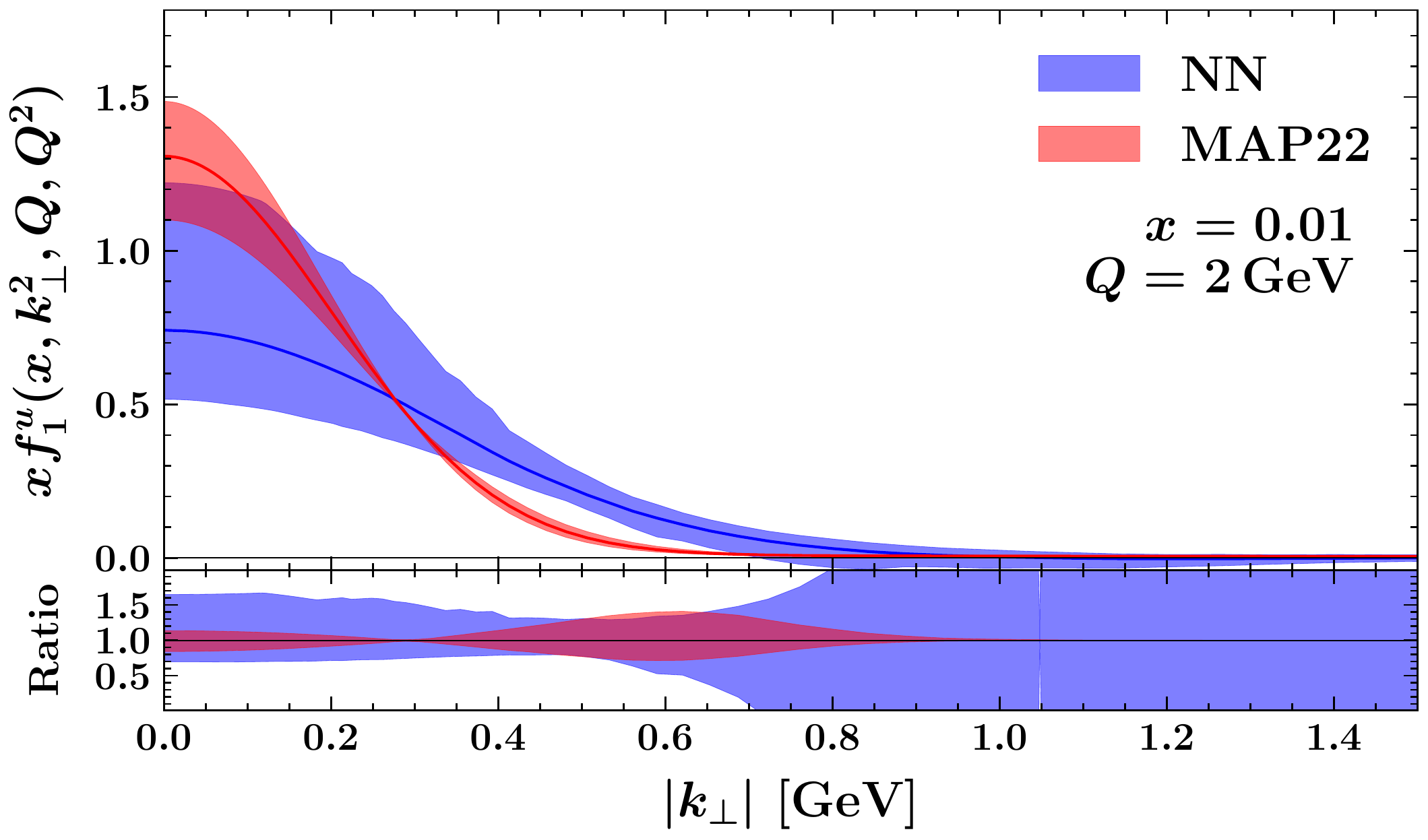}
    \caption{The unpolarised TMD PDF of the $u$-quark in the proton extracted
    using the NN (blue) and the MAP22 (red) parametrisations at $\mu =
    \sqrt{\zeta} = 2$ GeV and $x =0.01$ as functions of the quark transverse
    momentum $|\bm{k}_\perp|$. The upper panel shows the actual distributions,
    while the bottom panel shows their ratios to the respective central values.
    Error bands represent one-$\sigma$ uncertainties.}
    \label{f:tmds}
\end{figure}
In Fig.~\ref{f:tmds}, we show the unpolarised TMD PDF of the $u$-quark in the
proton at $\mu = \sqrt{\zeta} = 2$~GeV and $x=0.01$ as a function of the quark
transverse momentum $|\bm{k}_\perp|$. As before, blue and red bands correspond
to NN and MAP22, respectively. The upper panel displays the actual TMD
distributions, while in the lower panel they are normalised to the respective
central values. 

A generally good agreement between NN and MAP22 is observed, with the former
featuring a larger relative uncertainty band. This is a direct consequence of
the flexibility of the NN parametrisation. In this respect, it is interesting to
observe that the relative size of the NN uncertainty band remains fairly stable
up to $|\bm{k}_\perp| \sim 0.6$~GeV, while it tends to increase for larger
values of $|\bm{k}_\perp|$ because of the increasingly smaller central value. On
the contrary, the MAP22 relative uncertainty band shrinks as $|\bm{k}_\perp|$
increases. Moreover, it shows a node at intermediate values of $|\bm{k}_\perp|$.
This behaviour can be traced back to the rigidity of the parametrisation. 

The fact that the NN TMD PDF has larger uncertainties than the MAP22 one may
seem to contrast with the results shown in Fig.~\ref{f:obs_comp}, where cross
sections computed with the NN model display smaller uncertainties than MAP22.
This can be understood by noting that during the minimisation process the
greater flexibility of the NN model enables it to better adapt to the behaviour
of data. Consequently, the minimiser does not need to resort to large correlated
shifts to minimise the $\chi^2$, avoiding an inflation of the uncertainty in the
final outcome.

\subsection{Conclusions}\label{sec:conclusions}

In this work, we presented the first extraction of the unpolarised quark TMD
PDFs in the proton from a comprehensive set of DY data using a parametrisation
for the nonperturbative part based on a NN. Our results employ
state-of-the-art perturbative inputs reaching N$^3$LL accuracy, and leverage
modern numerical techniques, such as MC sampling for uncertainty propagation,
analytic computation of the gradient of the $\chi^2$ for a more accurate
exploration of the parameter space, cross-validation to avoid overfitting, and
full treatment of correlated experimental uncertainties.

By directly comparing the results obtained with the NN parametrisation to those
of the more traditional functional form of Ref.~\cite{Bacchetta:2022awv}
(MAP22), we showed that the former is superior in terms of fit quality. This
provides clear evidence that the NN-based parametrisation captures more
effectively the information encoded in the data than traditional
parametrisations. This is particularly evident for the ATLAS measurements, the
most precise data sets included in the fit. We showed that the flexibility of
the NN parametrisation allows us to better control the impact of correlated
uncertainties. For data sets characterised by large correlated systematics, such
as the fixed-target measurements, this typically leads to a significant
reduction of uncertainties.

This work serves as a proof of concept, opening new and exciting possibilities
for future high-precision and high-impact extractions. One of the main
advantages of NNs is that they scale particularly well with the complexity of
the task. More specifically, extending the data set to include semi-inclusive
deep-inelastic scattering data (which in turn requires a simultaneous fit of TMD
PDFs and TMD fragmentation functions) and introducing TMD flavour dependence is
a relatively straightforward task when using NNs. We are currently working along
these directions and plan to release a fully-fledged TMD extraction based on NNs
in the near future.

\subsection{Acknowledgments}

The work of S.R. is supported by the German Science Foundation (DFG), grant
number 409651613 (Research Unit FOR 2926), subproject 430915355. The work of
L.R. is partially supported by the Italian Ministero dell'Universit\`a e Ricerca
(MUR) through the research grant 20229KEFAM. This material is also based upon
work supported by the U.S. Department of Energy, Office of Science, Office of
Nuclear Physics under contract DE-AC05-06OR23177.
The contribution of C.B. is based upon work supported by Laboratory Directed Research and
Development (LDRD) funding from Argonne National Laboratory, provided by 
the Director, Office of Science, of the U.S. Department of Energy under 
Contract No. DE-AC02-06CH11357.

\bibliographystyle{myrevtex}
\bibliography{biblio}
\end{document}